\documentclass[conference]{IEEEtran}
\IEEEoverridecommandlockouts
\usepackage{amsmath,amssymb,amsfonts}
\usepackage{listings}
\usepackage[frozencache,cachedir=.]{minted}
\usepackage{algorithm}
\usepackage{algorithmicx}
\usepackage{nicematrix}
\usepackage[noend]{algpseudocode}
\usepackage{graphicx}
\usepackage{textcomp}
\usepackage{xcolor}
\usepackage{tikz}
\usepackage{makecell}
\usetikzlibrary{shapes,positioning,calc,arrows,chains}
\usepackage[unicode, hidelinks]{hyperref}
\usepackage[hyphenbreaks]{breakurl}
\usepackage[style=ieee, bibencoding=utf8, citestyle=numeric-comp]{biblatex}
\def\BibTeX{{\rm B\kern-.05em{\sc i\kern-.025em b}\kern-.08em
    T\kern-.1667em\lower.7ex\hbox{E}\kern-.125emX}}
\usepackage{booktabs}
\usepackage{colortbl}
\usepackage{multirow}
\usepackage{flushend}
\usepackage{verbatim}
\usepackage{subfig}

\algnewcommand{\algorithmicand}{\textbf{ and }}
\algnewcommand{\algorithmicor}{\textbf{ or }}
\algnewcommand{\OR}{\algorithmicor}
\algnewcommand{\AND}{\algorithmicand}

\definecolor{light-gray}{gray}{0.80}

\setminted{xleftmargin=3em}

\begin{document}

\title{LibAFL-DiFuzz: Advanced Architecture\\ Enabling Directed Fuzzing}

\author{
\IEEEauthorblockN{
  Darya Parygina\IEEEauthorrefmark{1}\IEEEauthorrefmark{2},
  Timofey Mezhuev\IEEEauthorrefmark{1}\IEEEauthorrefmark{2}, and
  Daniil Kuts\IEEEauthorrefmark{1}
}
\IEEEauthorblockA{
  \IEEEauthorrefmark{1}Ivannikov Institute for System Programming of the RAS
}
\IEEEauthorblockA{
  \IEEEauthorrefmark{2}Lomonosov Moscow State University
}
Moscow, Russia \\
\{pa\_darochek, mezhuevtp, kutz\}@ispras.ru
}

\maketitle


\begin{abstract}
Directed fuzzing performs best for targeted program testing via estimating the impact of each input in reaching predefined program points. But due to insufficient analysis of the program structure and lack of flexibility and configurability it can lose efficiency.

In this paper, we enhance directed fuzzing with context weights for graph nodes and resolve indirect edges during call graph construction. We construct flexible tool for directed fuzzing with components able to be easily combined with other techniques.
We implement proposed method in three separate modules: DiFuzzLLVM library for graph construction and indirect calls resolving, DiFuzz static analysis tool for processing program graphs and computing proximity metrics, and LibAFL-DiFuzz directed fuzzer based on LibAFL fuzzing library. We create additional LibAFL modules for enabling custom power scheduling and static instrumentation.
We evaluate indirect calls resolving and get increase in directed fuzzing efficiency for reaching deeper target points.
We evaluate context weights contribution and get benefits in TTE and scheduling iterations number.
We evaluate our fuzzer in comparison with AFLGo and BEACON, and reveal speedup in time to exposure on several benchmarks. Furthermore, our tool implements some important usability features that are not available in mentioned tools: target points detection, multiple target points support, etc.
\end{abstract}


\section{Introduction}

Software security is one of the main challenges in the modern world where technology plays a key role in all spheres of life. Coverage-guided fuzzing~\cite{aflpp,libfuzzer} becomes a prevailing dynamic testing approach in SSDLC~\cite{howard06,iso08,gost16}, which proved to be highly effective for detecting vulnerabilities. Directed fuzzing~\cite{aflgo}, on the other hand, is much better when it comes to the tasks of patch testing~\cite{katch}, crash reproduction~\cite{bugredux}, and static analysis report verification~\cite{christakis2016guiding}. To assess the helpfulness of the inputs in reaching a predefined program point fuzzer uses special proximity metrics.

Directed fuzzing tools generally use different graph representations of the target program to compute proximity metrics~\cite{aflgo,hawkeye}. The most effective approaches entail building sequences of some structural program components, such as basic blocks or functions, etc., to determine the impact of each input on reaching target points. These sequences are based on call graph and inter-procedural control-flow graphs of the program. The common problem of such methods lays in insufficient analysis of the graph structure. In the example~\ref{lst:example-main} we can see the widespread pattern of using indirect control transfers (indirect calls) that can pose a huge challenge for accurately constructing the call graph. Directed fuzzers usually utilize dynamic approaches for updating call graphs~\cite{parmesan,aflteam,dafl} but they require heavy-computation operations after each update, which affects performance. Our investigation revealed that indirect calls are mostly caused by virtual method calls. Therefore, we propose to use a static approach for hybrid fuzzing tools based on accurate handling of class hierarchies and their method calls.

\begin{listing}
\scriptsize
\begin{minted}[linenos=true,breaklines]{c++}
void foo() {
    if (x > 1) {
        cout << "branch 1T" << endl;
        if (x < 100)
            cout << "branch 2T" << endl;
        else
            cout << "branch 2F" << endl;
    } else {
        cout << "branch 1F" << endl;
    }
}

int main() {
    void (*ptr)(void) = &foo;
    *ptr();
    return 0;
}
\end{minted}
\caption{Example program with indirect call.}
\label{lst:example-main}
\end{listing}

Another key observation is that various target sequence components have a different impact on proximity metrics. As we see in the example~\ref{lst:example-main}, branch in line 2 is highly important for reaching line 9 whereas for reaching line 5 this branch is less important than branch in line 4. To address the issue of equal consideration of all the sequence components, we propose context weights-based proximity metrics. For each target sequence component, we determine its context weight as a combination of graph characteristics related to component position within the graphs.


This paper makes the following contributions:
\begin{itemize}
    \item We enhance directed fuzzing with context weights calculation for each target sequence component.
    \item We propose indirect edges resolving during call graph construction with the help of precise handling of class hierarchies and their method calls.
    \item We build directed fuzzing tool that consists of three modules: static analysis tool DiFuzz, graph construction library DiFuzzLLVM, and directed fuzzer LibAFL-DiFuzz based on LibAFL~\cite{libafl} fuzzing library. We evaluated our fuzzer in comparison with AFLGo~\cite{aflgo} and BEACON~\cite{beacon}, and revealed speedup in time to exposure on several benchmarks.
\end{itemize}

\section{Related Work}

\subsection{AFLGo: energy planning based on simulated annealing algorithm}
\label{sec_aflgo}

One of the first solutions in the field of directed greybox fuzzing was AFLGo~\cite{aflgo} tool. The authors solve the problem of exploring given target points in code by scheduling energy of the inputs based on simulated annealing algorithm~\cite{kirkpatrick1983optimization,vcerny1985thermodynamical}. According to the algorithm, during energy scheduling, large energy values are assigned to inputs that provide the closest approximation to the target points.

AFLGo treats the target point exploration problem as an optimization problem. The resulting set of inputs with the largest energy values converges asymptotically to a set of global optimal solutions that allow to explore the program execution paths closest to the target points. To regulate the convergence rate, the authors use the exponential law of decreasing parameter \textit{temperature}, which characterizes the probability of assigning large energy values to the worst solutions.

A metric based on the distances over the target basic blocks is proposed as a metric for the proximity of the program execution to reach the target points. Before fuzzing begins, a stage of static program instrumentation is performed: a program call graph and intra-procedural control flow graphs (CFGs) are constructed. Each vertex is matched with a metric value characterizing the distance from this vertex to all target vertices of the graph. The metric value for specific input is calculated as a normalized sum of metric values of each of the basic blocks of the program execution trace.
 support
The application of the metric proposed by the authors allows us to assign large energy values to those inputs that provide a shorter distance to the target points. However, such a solution has some drawbacks. First of all, due to the need for calculation metric values for all CFG basic blocks and all program functions, the stage of static instrumentation entails considerable time expenses and a large amount of unnecessary calculations. At that, any changes in the target program require the static instrumentation stage to be performed anew. In addition, the solution converges to the selection of input data providing the minimum path length to the target positions without considering the execution context, indirect transitions, and program data flow.

\subsection{Proximity metric based on target sequences}
\label{sec_leofuzz}

A popular approach to solving the problem of reaching target points is to use a metric based on sequences of some auxiliary program points --- so called target sequences. This approach provides the possibility of gradual selection of input data leading to the exploration of increasingly longer parts of the target sequences.

In LOLLY~\cite{lolly} tool, target sequences are completely specified by the user. A metric based on sequence coverage is used to calculate the proximity of program execution to reaching the target points. For specific input, the length of the longest common subsequence between the program execution trace and target sequences is calculated. The final value is arithmetic mean of this lengths for all target sequences.

Berry~\cite{berry} tool augments each user-defined target sequence with additional program points, constructing an enhanced target sequences (ETS). Berry uses inter-procedural control flow graph (iCFG) to choose those program points, that lay on any path reaching the sequence points (sequence point-dominator vertices). This approach allows to consider the context of target points. A proximity metric in Berry is build using the Knuth -- Morris -- Pratt~\cite{knuth1977fast} algorithm for finding the length of the longest matching prefix. The length of matching prefix with program execution trace is calculated separately for each ETS. The final metric value is also an arithmetic mean of all values calculated over each ETS.

LeoFuzz~\cite{leofuzz} also uses enhanced target sequences. ETSs are constructed using using the dominator trees for the program call graph and CFG of the specific function with target point. This method reduces the number of operations to compute the metric values and improves the efficiency of directed fuzzing. The proximity metric in LeoFuzz is computed based on the maximum length of a matching subsequence of the program execution trace on the given input and the ETSs. Initially, the metric value is determined separately for each ETS. The final metric value for the input is calculated based on the values of three metrics: the sequence coverage metric for the closest ETS, the priority metric for this ETS, and the maximum historical value of sequence coverage metric for this ETS. This approach allows multiple factors to be considered at once when scheduling the energy for the input, as well as separating inputs that provide best approximations for different ETSs. As a result, instead of solving the optimization problem for all target points at once, which can be extremely time consuming, optimization problems are solved separately for each of the target points.

All three tools use simulated annealing algorithm for energy scheduling. Berry and LeoFuzz also combine proximity metric with a coverage-guided metric to improve the efficiency of directed fuzzing.

\subsection{Directed fuzzing based on reachability analysis}

\subsubsection{BEACON}
\label{sec_beacon}

BEACON~\cite{beacon} uses reachability analysis to prune irrelevant basic blocks. It computes the reachability to target point for every basic block and then the irrelevant ones are pruned and they are not being instrumented. Interesting blocks are used in further analysis to apply constraints on them.

Then the backward interval analysis is used. It slices the program using statically computed control flow information and then applies the analysis that builds value intervals for all variables in program basic blocks. These value intervals are used to construct formulas for target reachability depending on value constraints. Then it inserts such formulas in code as instrumentation. At runtime values are checked for these constraints, if so, then program proceeds further to target point, otherwise it terminates.

\subsubsection{SieveFuzz}
\label{sec_sievefuzz}

SieveFuzz~\cite{sievefuzz} has a similar approach to BEACON, but with client-server architecture, where fuzzer sends information about target exploration to analysis component through LLVM pass. This component updates information about the dynamic control-flow graph.

This directed fuzzer has three main steps. The first, initial analysis, tries to determine whether the target is reachable (in current dynamic CFG). If not, it calls the second (exploration) step. The second step tries to resolve indirect calls and diversify the set of candidate seed traces with undirected fuzzing. At each step it updates dynamic CFG. When the target is reachable (after new CFG update), SieveFuzz calls the third step, called tripwired fuzzing. This step is running directed fuzzing, preempting and terminating execution of regions not within our target-reachable coverage set. It also collects information about indirect calls and updates the dynamic CFG.

\section{Directed fuzzing Pipeline}

We organize a directed fuzzing process into several standalone stages. This approach allows us to build a flexible and configurable analysis tool which could be easily re-implemented on different baseline frameworks. 

\begin{figure*}[ht]
\centering
\begin{center}
\includegraphics[scale=0.8]{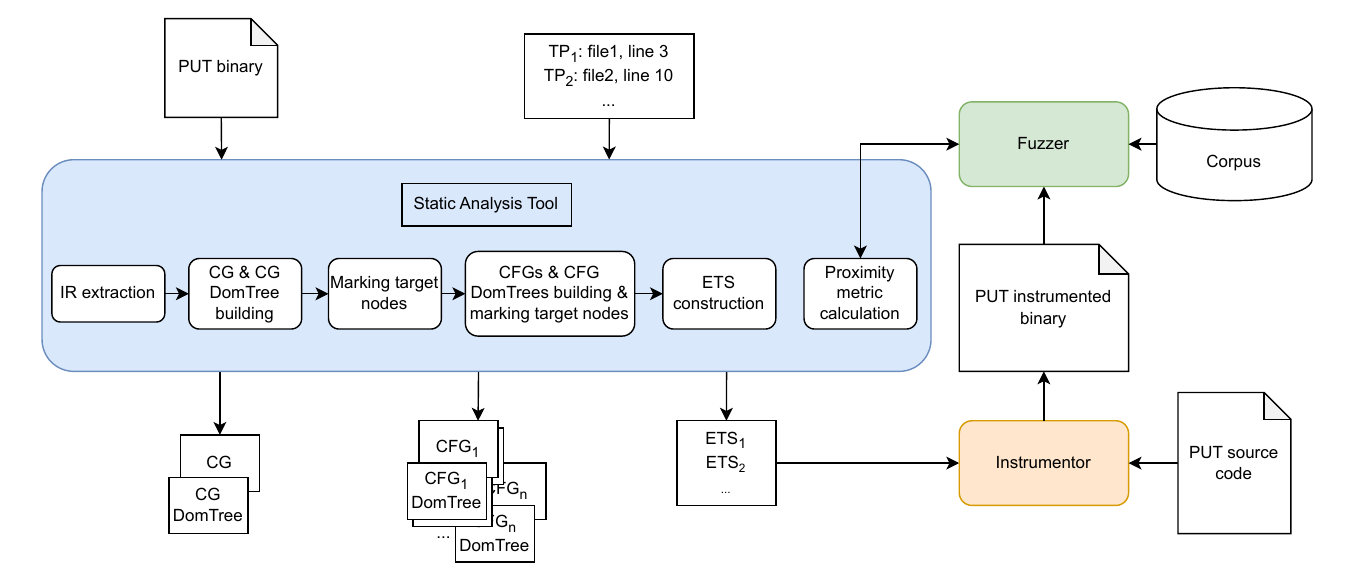}
\end{center}
\caption{Directed fuzzing pipeline.}
\label{fig:scheme}
\end{figure*}

Overall directed fuzzing pipeline can be illustrated by the scheme on fig.~\ref{fig:scheme}. The main instances implementing directed fuzzing stages are Static Analysis Tool, Instrumentor and Fuzzer. Static Analysis Tool performs preparation of the program under test (PUT) and collecting all the necessary information. Instrumentor provides a special instrumentation to the program, and Fuzzer performs actual directed fuzzing. During Fuzzer work, it communicates with Static Analysis Tool to calculate the proximity metric values for examined inputs.

\subsection{Static analysis stage}
\label{sec_static_analysis_stage}

Static Analysis Tool takes as input the binary of the PUT and the set of target points (TP) specified by source file name and line number. At the beginning, the Tool extracts PUT's intermediate representation for building graph representations of the program. Graphs are used to construct enhanced target sequences (ETSs), similar to LeoFuzz tool (section~\ref{sec_leofuzz}). First, the call graph of the program and its dominator tree are constructed. Then each target point of the given set is matched to corresponding call graph node in order to collect the set of target functions $S_{tf}$. Each function in $S_{tf}$ contains at least one target point. For each function $f \in S_{tf}$ inter-procedural control-flow graph $CFG_f$ and its dominator tree $DT_f$ are constructed. Such an optimization of selective CFG building allows to save time and resources avoiding construction CFGs for uninteresting functions. For each target point the Tool finds corresponding nodes in CFGs and marks them as targets, too.

After the dominator trees with marked target points are obtained, the next step is building ETS for each target point. ETS for target point $TP$ contains all the necessary nodes on the path from the beginning of the PUT to $TP$. For better reflecting dependencies between nodes and their relative importance in reaching target points each node is supplemented with its context weight value computed with respect to graph structural characteristics.

\subsection{Static instrumentation}
\label{sec:static_instrumentation}

When all the necessary information is obtained, PUT can be instrumented for providing ETS-related information while executing. More precisely, each basic block corresponding to graph node included in any ETS should denote the fact of its execution in runtime. The set of such basic blocks, actually executed while PUT running with some $input$, form ETS trace $T$ of the $input$. This trace is then used for calculating the values of proximity metrics for $input$.

\subsection{Directed fuzzing}

Once the program is statically instrumented, and all the necessary information is gathered, it is time to start fuzzing. The directivity of fuzzing is achieved through two principal actions: proximity metrics calculation and energy scheduling.

Proximity metrics denote how close some concrete $input$ is to reach target points when being fed to the PUT. They are calculated on every new input obtained after mutations, and these values are used to determine the capability of reaching target points. If the proximity metric value for $input$ is high then $input$ is likely to open new paths leading to target points.

The proximity metrics values are the basis for ETS power scheduling. The higher the value is the more energy is given to the input. Such decision is motivated by the fact that closer inputs are able to go through many branches on the path to some targets, and their mutation provides opportunities to go closer to the targets while preserving execution of the initial segments of the path. Therefore the number of mutations is calculated for each new input in accordance with the values of the proximity metrics.

All stages together form directed fuzzing pipeline. Static Analysis Tool and Instrumentor work before the fuzzing begins, produce the necessary information and files that are later sent to Fuzzer. During dynamic analysis stage, Fuzzer communicates to Static Analysis Tool that is able to compute proximity metrics values based on internally kept information.

\section{Static analysis}

The most of preparations for directed fuzzing is made by Static Analysis Tool. It takes binary of the PUT and the set of target points and produces PUT's graph representations, ETSs, calculates context weights. This section describes the details of graph building, indirect calls resolving, target points mapping, ETS construction, and context weights calculation.

\subsection{CG and CFG construction}

Based on the extracted LLVM IR~\cite{llvm_ir} of the PUT, the call graph (CG) containing only explicit call edges is built. This process is followed by indirect calls resolving and inserting new call graph edges (section~\ref{sec_indirect_calls}). After marking target nodes in call graph and creating $S_{tf}$ set (section~\ref{sec_static_analysis_stage}), inter-procedural control-flow graphs are built for each $f \in S_{tf}$. CFGs for other functions are not needed for further analysis as they do not affect ETS construction. Both CG and CFGs nodes must contain information about their locations in source code, and CG nodes must additionally contain corresponding function names. This information is later used for mapping target points to graph nodes and ETS construction.

\begin{figure}[h]
\centering
\includegraphics[scale=0.4]{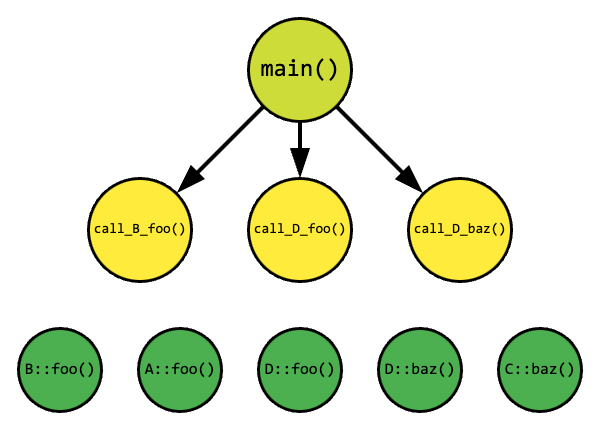}
\caption{Initial call graph with unresolved indirect calls for example~\ref{lst:example-hierarchy}.}
\label{fig:init-cg-hierarchy}
\end{figure}

Consider example listing~\ref{lst:example-hierarchy} with some class hierarchies in C++. After this stage the call graph for this example would look like in fig.~\ref{fig:init-cg-hierarchy} (for simplicity unimportant functions like constructor calls, \texttt{std::cout}, etc. are ignored).

\begin{listing}[h]
\scriptsize
\begin{minted}[linenos=true,breaklines]{c++}
class A {
public:
    virtual void foo() { cout << "A::foo\n"; }
};

class B : public A {
public:
    virtual void foo() { cout << "B::foo\n"; }
};

class C {
public:
    virtual void baz() { cout << "C::baz\n"; }
};

class D : public C, public A {
public:
    virtual void baz() { cout << "D::baz\n"; }

    virtual void foo() { cout << "D::foo\n"; }
};

void call_B_foo() {
    A *b = new B();
    b->foo();
    delete b;
}

void call_D_baz() {
    C *dc = new D();
    dc->baz();
    delete dc;
}

void call_D_foo() {
    A *da = new D();
    da->foo();
    delete da;
}

int main() {
    call_B_foo();
    call_D_baz();
    call_D_foo();
    return 0;
}
\end{minted}
\caption{Example with class hierarchy in C++.}
\label{lst:example-hierarchy}
\end{listing}

All implicit calls (\texttt{b->foo()}, \texttt{dc->baz()}, \texttt{da->foo()}) are missed after graph construction, because they cannot be resolved assuredly and so they are ignored. If some target point is located inside the corresponding methods, there is no way to find the path to it in such call graph.

\subsection{Indirect calls resolving}
\label{sec_indirect_calls}

\begin{figure}[h]
\centering
\includegraphics[scale=0.35]{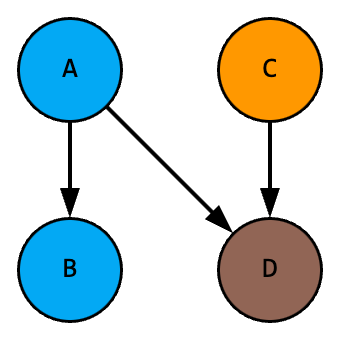}
\caption{Hierarchy tree for example~\ref{lst:example-hierarchy}.}
\label{fig:hierarchy-tree-hierarchy}
\end{figure}

This step is performed after CG construction to update it with some edges that are considered indirect calls. To do this, information about all class hierarchies (in case of C++ code), all overriden and virtual functions with their prototypes and names, and debug information of indirect calls is collected. At first, the hierarchy tree for each class hierarchy is built. The top of the tree is related to the base class, the leaves represent final classes, and there is an edge from node $n_1$ to node $n_2$ iff $n_2$ is directly derived from $n_1$. Fig.~\ref{fig:hierarchy-tree-hierarchy} illustrates the hierarchy tree built for example~\ref{lst:example-hierarchy}.

With the IR information we can determine the base class in the hierarchy for each indirect call, and take all the candidates from the members of the hierarchy tree with matched function name and prototype. In terms of the graphs, this technique allows us to resolve which CG nodes (functions) can be candidates for particular indirect calls, and add appropriate edges. We should notice that our method cannot lead to full indirect calls resolution as there is no enough information in LLVM IR, however it gives much better approximation than mapping only by function prototype.

For indirect calls that are not related to any hierarchy, such cases are considered as calls of function pointers, so for them we add all edges leading to nodes with matched function prototypes.

\begin{figure}[h]
\centering
\includegraphics[scale=0.35]{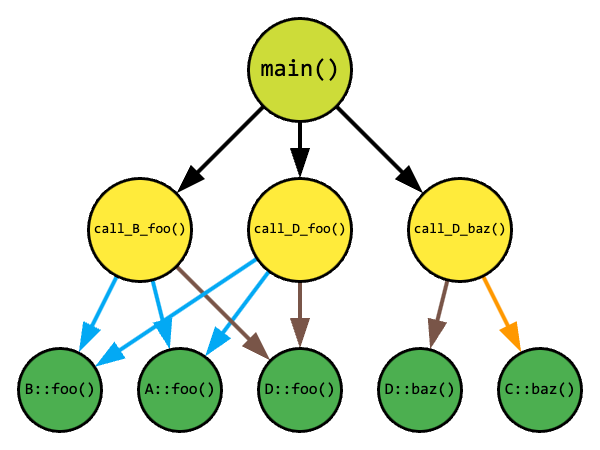}
\caption{Call graph with resolved indirect edges for example~\ref{lst:example-hierarchy}.}
\label{fig:cg-resolved-hierarchy}
\end{figure}

For the example~\ref{lst:example-hierarchy}, the resolved indirect calls are shown in the fig.~\ref{fig:cg-resolved-hierarchy}. We marked \texttt{A::foo}, \texttt{B::foo} and \texttt{D::foo} as candidates for \texttt{b->foo()} call because of \texttt{b} pointer membership in \texttt{A-B-D} hierarchy. As for \texttt{dc} pointer, we determined its relation to \texttt{C-D} hierarchy, and for \texttt{da} pointer we also added edges to functions from \texttt{A-B-D} hierarchy.

This example shows the basic behavior of our indirect call resolving algorithm: we rely on the hierarchy tree, base type of calling object, and name and prototype of called function.

\subsection{Mapping target point to graph nodes}
\label{sec_tp_mapping}

Target points are specified by source file name and line number. First of all, each target point should be mapped to exactly one node of PUT's call graph. The corresponding node should be related to the function which contains the target point. We consider that point $p$ is contained in the function $f$ iff source file name of $p$ matches file name with implementation of $f$, and the line number of $p$ is inside the borders of $f$ in the source code. When node $n$ corresponding to the target point $TP$ is found then $n$'s function name is added to $S_{tf}$ set.

After building CFGs for all selected functions, the similar procedure is performed on the nodes of CFGs. For each $TP$ we check nodes of the corresponding CFG to find the one which is related to the basic block involving the $TP$'s source line number. If there is no such basic block we try to map $TP$ to the closest basic block before or after $TP$ line number. Such decision is designed for the cases when the line in the source file doesn't contain program instructions, for example it may be empty line in the middle of the function body, or the function prototype.

\subsection{ETS construction}
\label{ets_construction}

The main concept of enhanced target sequences is that they accumulate all the necessary nodes on the paths from PUT's entry point to the target points. The idea of necessary nodes is perfectly expressed by domination property of the graph nodes~\cite{dragon_book}. The domination relation is defined conservatively as follows: node $m$ dominates node $n$ iff all the paths to node $n$ include node $m$. So instead of call graph and CFGs themselves, we use their dominator trees for ETS construction.

For each target point $TP_i$ with corresponding target nodes $n^{CG}_{TP_i}$ and $n^{CFG_i}_{TP_i}$ we add the node $n$ to $ETS_i$ iff it dominates either $n^{CG}_{TP_i}$ in call graph or $n^{CFG_i}_{TP_i}$ in $CFG_i$. So if we operate on dominator trees we need to collect two sequences of all the parents of $n^{CG}_{TP_i}$ in call graph and all the parents of $n^{CFG_i}_{TP_i}$ in $CFG_i$ recursively. Then we need to join them into one supersequence to form $ETS_i$. All the nodes added to ETSs are called intermediate ETS nodes.

\begin{figure}[h]
\centering
\includegraphics[scale=0.35]{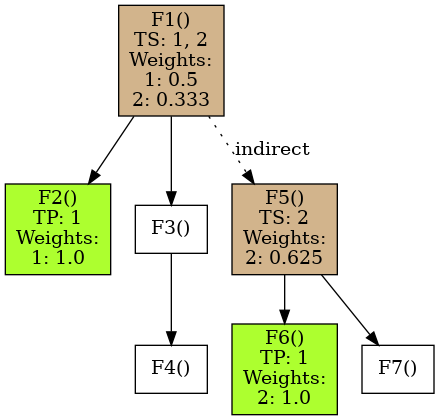}
\caption{Model call graph domtree with marked target points (green), intermediate ETS nodes (brown), and calculated context weights.}
\label{fig:model_cg_domtree}
\end{figure}

\begin{figure}[h]%
    \centering
    \subfloat[\centering CFG domtree for F2()]{{\includegraphics[scale=0.4]{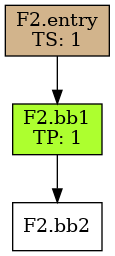} }
    \label{fig:model_cfgf2}}%
    \quad
    \subfloat[\centering CFG domtree for F6()]{{\includegraphics[scale=0.4]{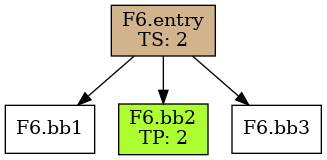} }
    \label{fig:model_cfgf6}}%
\caption{Model CFG domtrees with marked target points (green), intermediate ETS nodes (brown).}
\end{figure}

Consider the example call graph domtree in fig.~\ref{fig:model_cg_domtree} for some model program, and target functions CFGs domtrees in fig.~\ref{fig:model_cfgf2},~\ref{fig:model_cfgf6}. Let the target points be marked as green, then ETS intermediate nodes will be the ones marked as brown.

For the target point in function \texttt{F2()} ETS will include such nodes:

$$ETS_{F2} = \{ F1,\ F2,\ F2.Entry,\ F2.bb1\}.$$

For the target point in function F6():

$$ETS_{F6} = \{ F1,\ F5,\ F6,\ F6.Entry,\ F6.bb2\}.$$

\subsection{Weights calculating}

When ETSs for all targets are constructed, we can calculate context weights for CG domtree and CFG domtrees nodes. Weights are computed for each domtree independently using the structural characteristics of the domtree itself and its source graph. Each weight value correspond to one ETS, so if the node is related to $k$ ETSs then it will have $k$ weight values. If the node is not related to any ETS then it will not have any context weight values.

Let $distance_{ij}$ be the minimal distance from node $i$ to node $j$ in the graph, $level_i$ be the depth of $i$ in domtree (counting from 1), $rel\_succ_{it}$ -- the number of relevant successors for $i$ related to target point $t$ in the graph (i.e. successors for $i$ laying on some path to $t$), $succ_i$ -- the number of all successors for $i$ in the graph and $domsucc_i$ -- the number of successors for $i$ in domtree. For the distances calculation we apply the following rules:
\begin{enumerate}
    \item for $i$ and $j$ connected with edge $e$: $distance_{ij}$ = $distance_e$ = 2 if $e$ was added as indirect (section~\ref{sec_indirect_calls}), otherwise $distance_{ij}$ = $distance_e$ = 1;
    \item if there exists path from $i$ to $j$ then $distance_{ij} = min_P\sum_{e \in P} distance_e$ where $P$ iterates over the paths from $i$ to $j$;
    \item $distance_{ij}$ = 0 if there is no path from $i$ to $j$.
\end{enumerate}

 Using the introduced characteristics we can describe 4 metrics and the context weight formula:

$$Distance(i,\ t) = \frac{1}{distance_{it}}$$

$$Levels(i) = \frac{levels_i}{max\_levels}$$

$$Successors(i,\ t) = \frac{rel\_succ_{it}}{succ_i}$$

$$Probability(i) = P_i = \frac{1}{domsucc_i}$$
where $t$ is the target point node, $max\_levels$ -- the maximum depth in domtree.

And context weight metric is defined as follows:

$$weight_{ETS_k}(i) =$$
$$\frac{\left(Distance(i,\ t) + Levels(i) + Successors(i,\ t) + P_i\right)}{4}$$
where $i$ is the current node, $ETS_k$ -- ETS for context weight calculating, $t$ -- the target point node of $ETS_k$.

All of these metrics are defined in the range from 0 to 1 and all of them determine possible proximity to the target point by different parameters. For the target points themselves the value of weight is not calculated and set to 1. For the intermediate ETS nodes context weight metric is constructed in such a way that when approaching the target point the $weight$ value increases to 1.

Consider the above example with call graph domtree shown in fig.~\ref{fig:model_cg_domtree}. Node \texttt{F1} is included in both $ETS_1$ and $ETS_2$, so it has the value of weight for each of them. Nodes \texttt{F5} and \texttt{F6} belong to $ETS_2$ only, whereas \texttt{F2} belongs to $ETS_1$. Their weight values are related to the only ETS they correspond.

\begin{figure}[h]
\centering
\includegraphics[scale=0.35]{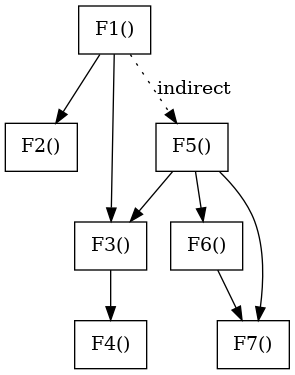}
\caption{Model call graph.}
\label{fig:model_cg}
\end{figure}

Let the original call graph for above example be shown in fig.~\ref{fig:model_cg}. Then context weight value for \texttt{F1} node for $ETS_2$ will be calculated in the following way: $distance_{F1F6}$ = 2 + 1 = 3 because of indirect edge \texttt{F1 -> F5}; $level_{F1}$ = 1, $max\_levels$ = 3; $rel\_succ_{F1F6}$ = 1, $succ_{F1}$ = 3; $domsucc_{F1}$ = 3.

$$weight_{ETS_2}(F1) = \frac{1}{4}\cdot \left(\frac{1}{3} + \frac{1}{3} + \frac{1}{3} + \frac{1}{3}\right) \approx 0.333$$

Another $weight$ values are computed in the same way.

\section{Directed fuzzing}

Our method proposes two main techniques that determine the direction of fuzzing: proximity metrics and ETS-based power scheduling. This section describes their work in detail, and shows Fuzzer architecture components that are necessary for them to be implemented.

\begin{figure}[h]%
    \centering
    \subfloat[\centering Call graph domtree for example~\ref{lst:example-main}]{{\includegraphics[scale=0.35]{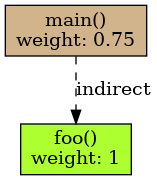} }
    \label{fig:main_cg}}%
    \quad
    \subfloat[\centering CFG domtree for \texttt{foo()} for example~\ref{lst:example-main}]{{\includegraphics[scale=0.35]{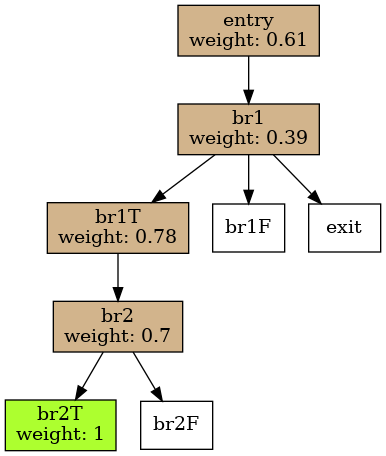} }
    \label{fig:main_cfg_foo}}%
\caption{Domtrees for example~\ref{lst:example-main}.}
\end{figure}

In order to demonstrate the application of proximity metrics, fuzzing instances, and power scheduling we will use example~\ref{lst:example-main}. Let the target point be line 5, then the call graph domtree and CFG for function \texttt{foo} with marked ETS are shown in fig.~\ref{fig:main_cg},~\ref{fig:main_cfg_foo}.

\subsection{Fuzzing instances for ETS power scheduling}

Several important Fuzzer components must be implemented to organize directed fuzzing. Their interaction will provide information transfer between the PUT, Fuzzer and Static Analysis Tool.

The first such component is an Observer instance. In general, Observer receives some characteristics of the PUT execution with some input and selects and saves the data needed for other Fuzzer components. In our case, ETS Observer is responsible for selecting the ETS trace $T$ (section~\ref{sec:static_instrumentation}) from each PUT execution and saving it alongside the corresponding input. For example~\ref{lst:example-main}, possible ETS trace may be the following: $T = \{main,\ foo,\ entry,\ br1,\ br1F,\ exit\}$.

After the PUT execution the corresponding input should be evaluated as interesting or not. Interesting inputs are added to the corpus and used for further mutations. Interestingness of the input is usually determined by its novelty value for some criteria. In case of coverage-based fuzzing, the novelty often means examining new edges in the coverage map~\cite{afl_coverage_map}. In accordance to ETSs, we consider novelty to be a new unique ETS trace that was never seen on previous inputs. This allows us to explore new paths that may lead to new code locations and may help reaching target points, and avoid adding to the corpus many inputs with the same ETS trace that may made fuzzing stuck in the same paths.

The Fuzzer component responsible for determining interestingness of the input is called Feedback. It takes information about execution characteristics from the Observer and processes them trying to detect novelty. Corresponding input is added to the corpus if Feedback result is positive. In our method we use the combined Feedback that implements logical OR above ETS Feedback and Map Feedback (which tracks opening new edges in the coverage map) in order to monitor not only reaching target points but also the overall PUT coverage. The essence of such decision is that exploring new paths can significantly improve the ability of reaching target points because of greater chances of reaching some intermediate ETS nodes.

The main component responsible for providing direction of fuzzing process is the instance of PowerScheduler. In general, power scheduling is the process of assigning inputs with the natural number from fixed range. This number denotes the amount of times the input is mutated and evaluated with the PUT. The greater power value is the more different variants Fuzzer can obtain from the input. ETS PowerScheduler receives ETS trace from corresponding Observer for some input from the corpus and calculates energy applying power schedule scheme described in section~\ref{sec_power_schedule_scheme}.

\begin{figure*}[h]
\centering
\includegraphics[scale=0.8]{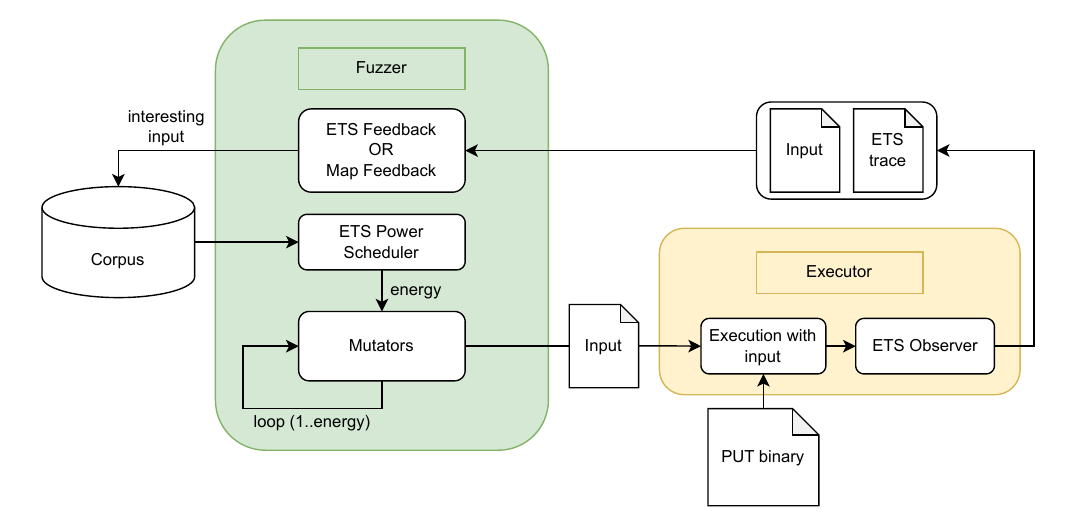}
\caption{Scheme of Fuzzer components for ETS-based power scheduling.}
\label{fig:fuzzer-components}
\end{figure*}

Fig.~\ref{fig:fuzzer-components} illustrates interactions of the main Fuzzer components and data transfers between them.

\subsection{Proximity metrics}

When the program ETS trace is delivered to ETS Feedback, the following proximity metrics are needed to be calculated:
\begin{itemize}
  \item \textit{PriorityW}
  \item \textit{SeqCovW}
  \item \textit{gMaxCovW}
\end{itemize}

These metrics are inspired by LeoFuzz approach (section~\ref{sec_leofuzz}), but unlike it we use the calculated $weight$ values instead of just counting the number of elements in sequences.

\textit{PriorityW} metric shows the priority of every ETS for ETS PowerScheduler. $PriorityW_i$ means degree of similarity between $ETS_i$ and all other ETSs. If one ETS is as similar as possible to the others then it's easier to reproduce. The metric is evaluated with the following formula:

$$PriorityW_i = $$
$$\sum\limits_{\substack{j=1,\\ j \neq i}}^{N}\left(\frac{SIMW(ETS_i, ETS_j)}{max(W(ETS_i, ETS_j), W(ETS_j, ETS_i))} \geq \epsilon?1:0\right)$$
where

$$SIMW(S_i, S_j)=$$
$$\max_{C=CS(S_i, S_j)} \sum_{k=1}^{len(C)} \frac{weight_{S_i}(C_k) + weight_{S_j}(C_k)}{2}$$
$C=CS(S_i, S_j)$ means iterate over all common subsequences for $S_i$ and $S_j$; if one of $S_i$ and $S_j$ is ETS trace then the corresponding $weight$ values are considered equal to $weight$ values of another sequence; $\epsilon$ is calculated empirically (we take the geometric mean of $PriorityW$ values);

$$W(S_i, S_j)=SIMW(S_i, S_j) + \sum_{n \in S_i \backslash C_{max}} \frac{1}{weight_{S_i}(n)}$$
-- summed inversed contribution of sequence $S_i$ to the similarity of $S_i$ and $S_j$; $C_{max}$ -- such common subsequence for $S_i$ and $S_j$ that gives the maximum value of $SIMW$. The lower value of $W$ means the greater contribution to similarity.

\textit{SeqCovW} metric shows similarity between program ETS trace $T$ and some $ETS_i$. The larger the value of this metric is the more likely mutating the corresponding input will reach $TP_i$. The following formula is used to calculate \textit{SeqCovW}:

$$SeqCovW_i(ETS_i, T) = \frac{SIMW(ETS_i, T)}{W(ETS_i, T)}$$

The last \textit{gMaxCovW} metric is evaluated and updated during fuzzing runtime. This value is an indication of how much particular ETS has been covered by the current corpus. It is updated when ETS trace is already processed (\textit{SeqCovW} for this trace is calculated). The formula for this metric is:
$$gMaxCovW_i = max(gMaxCovW_{i\_old}, SeqCovW_i(ETS_i, T)).$$

When new $input$ is added to the corpus during fuzzing, the calculations of \textit{SeqCovW} and \textit{gMaxCovW} metrics are performed for each ETS. Then ETS corresponding the highest \textit{SeqCovW} value is chosen as input's Outstanding Target Sequence ($OTS$), similar to LeoFuzz approach. Based on the metrics values for this $OTS$ we can then compute the overall impact of the $input$ on reaching $OTS$-related target point.

To mitigate the difficulties in estimating \textit{gMaxCovW} values when the total program coverage is small, we apply LeoFuzz decision and use a special Weighted Comprehensive Factor ($CFW$) that displays the compound metrics value with the special handling of \textit{gMaxCovW}. When computing $CFW$, we use $OTS$ $PriorityW$ and $OTS$ $SeqCovW$ values, taking into account $OTS$ $gMaxCovW$ only if more than the half ETSs have \textit{gMaxCovW} exceeding $\beta$ threshold value (in our case $\beta = 0.5$). The resulting $CFW$ is calculated by the following formulae:

$$CFW = \begin{cases} \frac{1}{2} \left(SeqCovW + \frac{PriorityW}{N}\right), \\ \phantom{-}\mbox{if } \sum_{j=1}^{N}\left(gMaxCovW_j \geq \beta~?~1 : 0\right) < \frac{N}{2} \\ \frac{1}{3} \left(SeqCovW + \frac{PriorityW}{N} + (1 - gMaxCovW)\right), \\ \phantom{-}\mbox{if } \sum_{j=1}^{N}\left(gMaxCovW_j \geq \beta~?~1 : 0\right) \geq \frac{N}{2} \end{cases}$$
where N is the number of ETSs.

\begin{figure}[h]
\centering
\includegraphics[scale=0.25]{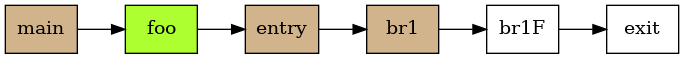}
\caption{ETS trace $T_1$ for example~\ref{lst:example-main}.}
\label{fig:main-ets1}
\end{figure}

\begin{figure}[h]
\centering
\includegraphics[scale=0.25]{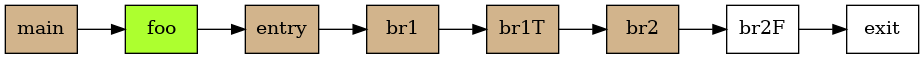}
\caption{ETS trace $T_2$ for example~\ref{lst:example-main}.}
\label{fig:main-ets2}
\end{figure}

Consider the following ETS traces $T_1$, $T_2$ (fig.~\ref{fig:main-ets1},~\ref{fig:main-ets2}) related to example~\ref{lst:example-main}. ETS for the only target point in line 5 is: $ETS_1 = \{main,\ foo,\ entry,\ br1,\ br1T,\ br2,\ br2T\}$. Context weights for this ETS are shown inside nodes of call graph domtree and CFG domtree for function foo in fig.~\ref{fig:main_cg},~\ref{fig:main_cfg_foo}.

Then the proximity metrics will be calculated in the following way:

$$SIMW(ETS_1, T_1) = 0.75 + 1 + 0.61 + 0.39 = 2.75,$$
$$SIMW(ETS_1, T_2) = 0.75 + 1 + 0.61 + 0.39 + 0.78 + 0.7 = 4.23,$$
$$W(ETS_1, T_1) = SIMW(ETS_1, T_1) + \frac{1}{weight_{ETS_1}(br1T)} +$$
$$+ \frac{1}{weight_{ETS_1}(br2)} + \frac{1}{weight_{ETS_1}(br2T)} =$$
$$= 2.75 + \frac{1}{0.78} + \frac{1}{0.7} + \frac{1}{1} \approx 6.46,$$
$$W(ETS_1, T_2) = SIMW(ETS_1, T_2) + \frac{1}{weight_{ETS_1}(br2T)} =$$
$$= 4.23 + \frac{1}{1} = 5.23,$$
$$PriorityW_1 = 0,$$
$$SeqCovW_1(ETS_1, T_1) = \frac{2.75}{6.46} \approx 0.43,$$
$$SeqCovW_1(ETS_1, T_2) = \frac{4.23}{5.23} \approx 0.81.$$

Supposing \textit{gMaxCovW} value has not reached threshold value, then $CFW$ values for $T_1$ and $T_2$ equal corresponding \textit{SeqCovW} values divided by 2, so we can say that the input related to $T_2$ has more impact on reaching the target point.

\subsection{Power schedule scheme}
\label{sec_power_schedule_scheme}

The direction of fuzzing is mostly guided by power scheduling technique. In our method we apply simulated annealing algorithm proposed by AFLGo (section~\ref{sec_aflgo}). This algorithm is proved to be a good solution for finding global optimum with the help of $temperature$ control parameter. Parameter defines the probability of accepting worse solution depending on passed time. This probability allows to get out of local optimums if there are ones. With the time passed the $temperature$ decreases, so does the probability.

We consider the optimum to be a solution that allows to reach the maximum value of $CFW$. The temperature is cooled from the initial value of $T_0 = 1$ according to the exponential law:

$$T = \alpha^k \cdot T_0$$
with $0.8 \leq \alpha \leq 0.99$. With the time budget limited there is a need to specify time $t_x$ spent to exploration phase (i.e. finding new various inputs) before switching to exploitation (i.e. aiming to reach target points). As in AFLGo, this switch is made when $T_k < 0.05$. So to compute the temperature $T$ at the given time $t$ we need to make some calculations:

$$T_k = 0.05 = \alpha^{k_x} \Rightarrow k_x = \frac{log(0.05)}{log(\alpha)}$$

$$k = \frac{t}{t_x}\cdot k_x$$

$$T = \alpha^k \cdot T_0 = \alpha^{\frac{t}{t_x}\cdot \frac{log(0.05)}{log(\alpha)}} = 20^{-\frac{t}{t_x}}$$

Using this relation we can calculate the value of $T$ at any given moment $t$. For energy scheduling we combine standard $energy$ value provided by fuzzer's power schedule with the value of $capabilityW$ computed by the following formula:

$$capabilityW = CFW \cdot (1 - T) + 0.5 \cdot T$$

$capabilityW$ describes degree of input's ability to reach its $OTS$ (the closest ETS). As time passes and $t$ decreases, the $CFW$ value begins to play bigger role. Finally, the overall value of energy is computed using the following relation:

$$ETSenergy = energy \cdot 2^{(capabilityW - 0.2) \cdot 10}$$

\section{Implementation}

We implemented proposed method in three separate modules. We wrote the library in C++ for graph construction and indirect calls resolving. We created the static analysis tool in Rust that makes all the preparation work and saves necessary information, and in addition computes proximity metrics values via the interface functions. We added several modules to LibAFL~\cite{libafl} fuzzing library written in Rust and the LLVM compiler pass~\cite{llvm} for ETS-related instrumentation, and built our directed fuzzer using added modules.

The modular structure allows our tool to have a number of important features.
It is very convinient to configure fuzzer to suite the ones needs as it allows to combine
different fuzzing algorithms with proposed directed fuzzing method. Within LibAFL library we implemented gathering detailed statistics for directed fuzzing and reporting on reaching target points. Other directed fuzzers don't have such functionaluty and can detect reaching target point only if it is related to crash, which drastically reduces their potential. Another important feature is multi-targeted directed fuzzing (not supported by many other tools).

In the next parts of this section we discuss implementation details related to each of these modules.

\subsection{DiFuzzLLVM}
\label{sec_impl_difuzz_llvm}

DiFuzzLLVM is our library that is used for CG and CFGs constructing. It also provides indirect calls resolving stage (described in~\ref{sec_indirect_calls}) and debug information extracting for graph's nodes.

We require PUT binary to be compiled with \texttt{wllvm}~\cite{wllvm} compiler that extracts LLVM bitcode (.bc format~\cite{llvm_bitcode}) modules and adds them to the binary. 
In DiFuzzLLVM we use LLVM-18 API to analyze extracted bitcode, create call graph and required CFGs.
We also extend graphs with debug attributes (source file name, line number, etc.) and dump them into DOT~\cite{dot} file format.



Indirect calls resolving takes place during call graph constructon.
At first we collect information about class inheritance links from LLVM metadata nodes that store information about base and derived classes.
Using information from these nodes we build hierarchy tree like in example~\ref{lst:example-hierarchy}. In the next step we find all functions from CG that relate to some class from hierarchy tree and save them for every hierarchy that contains this class.
At last we find all indirect call instructions (using \texttt{llvm::CallBase}), and for each of them we try to find the base class of the call source. Then we add edges to saved functions with matching type and name for every hierarchy the base class is contained in. When graph is dumped, we add \texttt{[indirect]} attribute for indirect call edges for later processing.

\subsection{DiFuzz}
\label{sec_impl_difuzz}

DiFuzz is our tool that implements the static analysis stage described in section~\ref{sec_static_analysis_stage}. DiFuzz takes as inputs precompiled binary (same as for DiFuzzLLVM) and the set of target points configured via TOML~\cite{toml} file as source file and line number pairs.

PUT call graph and CFGs are constructed via DiFuzzLLVM library, and saved to DOT files. Dominator trees for all graphs are built based on these DOT files with the help of \texttt{domtree}~\cite{domtree} crate, and are also saved to DOT files. Additional attributes (section~\ref{sec_impl_difuzz_llvm}) are preserved via parsing and dumping DOTs. For example, DOT file for call graph of a simple example program can look like in listing~\ref{lst_dot}.

\begin{listing}
\scriptsize
\begin{minted}[breaklines]{dot}
digraph "Call graph: example.ll" {
    label="Call graph: example.ll";

    Node0x10 [shape=record, filename="/home/user/example/main.c", startline=5,endline=10,label="{foo}"];
    Node0x20 [shape=record, filename="/home/user/example/main.c", startline=12,endline=23,label="{main}"];
    Node0x20 -> Node0x10;
}
\end{minted}
\caption{DOT file for example call graph.}
\label{lst_dot}
\end{listing}

Mapping between target points and graph nodes is made based on info from config and additional DOT attributes.

After we construct dominator trees for CG and CFGs and parse target points, we build ETS. We map call graph nodes to the entry BBs of corresponding functions. Every ETS member is marked in graph node structure in a special field mapping ETS ID to member type. ETS construction algorithm is presented in section~\ref{ets_construction}. In the end ETSs and calculated context weights are dumped to DOT files as parts of nodes' attributes. The example of DOT node is shown in listing~\ref{lst:node-weights}. \texttt{ts\_numbers} stores tuple of ETS IDs, \texttt{ts\_kind} contains tuple of ETS member types (target point or intermediate member) in corresponding ETSs, \texttt{w\_numbers} and \texttt{w\_val} attributes represent ETS IDs and corresponding weight values.

\begin{listing}
\scriptsize
\begin{minted}[breaklines]{dot}
Node0x30 [shape=record, filename="/home/user/example/main.cpp", startline=52,endline=77,startcolumn=9, label="{main}", ts_kind="point/member",ts_numbers="0/1", w_val="1.0/0.435",w_numbers="0/1"];
\end{minted}
\caption{Example DOT Node with ETS and weights inserted.}
\label{lst:node-weights}
\end{listing}

For the convenience of communicating with LibAFL, we create a TOML configuration file \texttt{"ets.toml"}, in which we store debug info for the graph nodes that are part of any ETS, and for ETSs themselves, together with the sequence of their weights and the calculated $PriorityW$ metric values.

\subsection{Static instrumentation}

To perform ETS-related instrumentation we need to know which basic blocks should be instrumented. They are basic blocks included in at least one ETS, and their debug information is saved in the \texttt{"ets.toml"} file (section~\ref{sec_impl_difuzz}).

Instrumentation is performed during PUT compilation via specially implemented LLVM-pass. Inside the pass, for each program module we are iterating over all the functions in this module and over all the basic blocks of each function. For each basic block we find the first instruction and extract the source line number and column number from its LLVM DbgInfo. The goal is to find the first meaningful location inside the basic block body, so if two instructions have same line numbers but different column numbers then we choose a minimal one.

We add LLVM instructions into LLVM IR file to the beginning of the matching basic blocks with required graph nodes from \texttt{"ets.toml"} to make the PUT print basic blocks ID into the special shared memory area while PUT execution.


\subsection{Fuzzing}


We implemented ETS Observer, ETS Feedback, and ETS PowerScheduler as corresponding LibAFL modules. Components interact with the help of special Metadata instances. Each instance is connected to some component. We have one Metadata instance for ETS Observer that holds ETS trace of the current PUT execution as a set of basic block IDs. Another Metadata is connected to ETS Feedback, and it holds path to \texttt{"ets.toml"} file, vectors of \textit{PriorityW} and \textit{gMaxCovW} values for each ETS, and prefix trie implemented with \texttt{trie\_rs}~\cite{trie_rs} crate to hold the variety of ETS traces seen on the previous PUT executions. We have Metadata instance related to Testcase: it holds the values of $CFW$ and $capabilityW$ for the input of the current PUT execution so these data may be used anywhere the input is available.



The resulting fuzzer LibAFL-DiFuzz is constructed based on standard LibAFL canvas for AFL-like fuzzer with addition of ETS components.

\section{Evaluation}

For evaluating our method we estimated the following metrics: Time to Exposure (TTE) and the total number of found crashes within the limited time. TTE shows how much time the instrument takes to find specific error. The total number of crashes allows to estimate the average speed of crash finding. In a separate experiment we evaluated context weights contribution to our method with the help of two metrics: TTE and scheduling iterations number.

Experimental setup included one machine with Intel(R) Core(TM) i7-9700 CPU, 32 GiB RAM and Ubuntu 22.04 LTS, and another machine with two AMD EPYC 7542 32-Core CPUs, 512 GiB RAM and Ubuntu 20.04 LTS. We chose two instruments to compare our prototype with: AFLGo~\cite{aflgo} and BEACON~\cite{beacon}. AFLGo is the first and widely popular directed greybox fuzzer, BEACON proposes ideas of pruning irrelevant basic blocks with reachability analysis (section~\ref{sec_beacon}).
For TTE estimation each instrument was launched with $t_x$ = 120 minutes for power scheduling, and the time from start to reporting specific crash was recorded. For estimating the total number of crashes we launched each fuzzer with specified timeout and recorded the total number of files after time expiration. We conducted each experiment several times and took the best attempt filtering outliers caused by randomness effect.

As for the targets, we used examples from AFLGo -- \texttt{cxxfilt}, \texttt{giflib}, \texttt{jasper}, \texttt{objdump} and \texttt{libxml} programs of versions 2.26.20160125, 5.1.0, 1.900.1, 2.28.51.20170419 and 2.9.2, respectively. For each program we launched it with AFLGo and picked some crashes that are discoverable within adequate amount of time (less than 60 minutes) and set corresponding target points. Table~\ref{tbl:targets} shows locations of target points, related IDs, and timeout for running experiments. For the total crash number estimation we took sets of target points from AFLGo corresponding examples with several extra points.

\begin{table}[htbp]
    \begin{center}
    \small
    \begin{NiceTabular}{c | >{\columncolor[gray]{0.9}}c | c | c}
        \hline
        \textbf{Project}&\textbf{TP ID}&\textbf{Location}&\textbf{TimeOut}\\
        \hline \hline

        \multirow[c]{4}{*}[0pt]{cxxfilt} & cxxfilt$_1$ & binutils/cxxfilt.c: 62 & \multirow[c]{4}{*}[0pt]{45 minutes} \\
        \cline{2-3}
        & cxxfilt$_2$ & libiberty/cplus-dem.c: 1489 & \\
        \cline{2-3}
        & cxxfilt$_3$ & libiberty/cp-demangle.c: 1596 & \\
        \cline{2-3}
        & cxxfilt$_4$ & libiberty/cplus-dem.c: 4319 & \\
        \hline
        
        \multirow[c]{2}{*}[0pt]{giflib} & giflib$_1$ & util/gifsponge.c: 61 & \multirow[c]{2}{*}[0pt]{30 minutes} \\
        \cline{2-3}
        & giflib$_2$ & util/gifsponge.c: 76 & \\
        \hline
        
        jasper & jasper$_1$ & \Block[]{}{src/libjasper/\\mif/mif\_cod.c:491} & 30 minutes \\
        \hline

        \multirow[c]{3}{*}[0pt]{objdump} & objdump$_1$ & objdump.c: 2435 & \multirow[c]{2}{*}[0pt]{30 minutes} \\
        \cline{2-3}
        & objdump$_2$ & objdump.c: 3589 & \\
        \cline{2-3}
        & objdump$_3$ & objdump.c: 3678 & \\
        \hline
        
        libxml & libxml$_1$ & valid.c:1189 & 60 minutes \\
        \hline
    \end{NiceTabular}
    \caption{Target points in example programs used for experiments.}
    \label{tbl:targets}
    \end{center}
\end{table}

\subsection{Indirect calls}

For evaluating indirect calls resolving we created synthetic example program with class hierarchy and virtual method calls. We marked 5 target points in this example within virtual methods and sorted them from the easiest to reach (\textit{hierarchy}$_1$) to the hardest (deepest in call graph -- \textit{hierarchy}$_5$).

Table~\ref{tbl:exp_icalls} shows the results of evaluating indirect calls resolving. Columns \textbf{Icalls} and \textbf{No icalls} denote two experimental LibAFL-DiFuzz setups: with indirect calls resolving enabled, and without it. Columns \textbf{TTE} show time to exposure for specified target points, and columns \textbf{Iters} mean scheduling iterations number made by LibAFL-DiFuzz before the target point was discovered.

According to the table, in most cases indirect calls resolving helps to reach target points faster and reduce the number of scheduling iterations. What is more important, the significant growth in directed fuzzing efficiency is related to the hardest to reach target points.

\begin{table}[htbp]
    \begin{center}
    \small
    \begin{NiceTabular}{c | c | >{\columncolor[gray]{0.9}}c | c | >{\columncolor[gray]{0.9}}c | c}
        \hline
        \multirow[c]{2}{*}[0pt]{\textbf{№}} & \multirow[c]{2}{*}[0pt]{\textbf{TP ID}} &  \multicolumn{2}{c}{\textbf{Icalls}} & \multicolumn{2}{c}{\textbf{No icalls}} \\
        \cline{3-6}
        & & \cellcolor[gray]{0.9}\textbf{TTE} & \textbf{Iters} & \cellcolor[gray]{0.9}\textbf{TTE} & \textbf{Iters} \\
        \hline\hline
        1 & hierarchy$_1$ & \cellcolor[gray]{0.9}11.69 & 6 & \cellcolor[gray]{0.9}\textbf{1.73} & 1 \\
        \hline
        2 & hierarchy$_2$ & \cellcolor[gray]{0.9}11.57 & 6 & \cellcolor[gray]{0.9}\textbf{5.30} & 3  \\
        \hline
        3 & hierarchy$_3$ & \cellcolor[gray]{0.9}\textbf{19.56} & 21 & \cellcolor[gray]{0.9}171.63 & 238 \\
        \hline
        4 & hierarchy$_4$ & \cellcolor[gray]{0.9}\textbf{19.62} & 21 & \cellcolor[gray]{0.9}102.96 & 120 \\
        \hline
        5 & hierarchy$_5$ & \cellcolor[gray]{0.9}\textbf{15.77} & 21 & \cellcolor[gray]{0.9}102.06 & 120 \\
        \hline
    \end{NiceTabular}
    \caption{Indirect calls resolving experiments.}
    \label{tbl:exp_icalls}
    \end{center}
\end{table}

\subsection{Context weights}

For evaluating context weights contribution we launched project examples from AFLGo (table~\ref{tbl:targets}) for the specified amount of time. Table~\ref{tbl:exp_weights} shows the results of evaluating context weights contribution. Columns \textbf{Weights} and \textbf{No weights} denote two experimental LibAFL-DiFuzz setups: with context weights enabled, and without them (that means comparisons between execution traces and ETSs based on longest common subsequence values). Columns \textbf{TTE} show time to exposure for specified target points, and columns \textbf{Iters} mean scheduling iterations number made by LibAFL-DiFuzz before the target point was discovered.

The results show that context weights increase directed fuzzing efficiency for all target points. Both TTE and number of scheduling iterations decrease when weights are enabled. The difference in TTE reaches 159 seconds for \textit{cxxfilt$_3$}, and iterations number distinction grows up to 77 for \textit{jasper$_1$}.

\begin{table}[htbp]
    \begin{center}
    \small
    \begin{NiceTabular}{c | c | >{\columncolor[gray]{0.9}}c | c | >{\columncolor[gray]{0.9}}c | c}
        \hline
        \multirow[c]{2}{*}[0pt]{\textbf{№}} & \multirow[c]{2}{*}[0pt]{\textbf{TP ID}} &  \multicolumn{2}{c}{\textbf{Weights}} & \multicolumn{2}{c}{\textbf{No weights}} \\
        \cline{3-6}
        & & \cellcolor[gray]{0.9}\textbf{TTE} & \textbf{Iters} & \cellcolor[gray]{0.9}\textbf{TTE} & \textbf{Iters} \\
        \hline\hline
        1 & cxxfilt$_1$ & \cellcolor[gray]{0.9}\textbf{0.53} & 1 & \cellcolor[gray]{0.9}1.21 & 1 \\
        \hline
        2 & cxxfilt$_2$ & \cellcolor[gray]{0.9}\textbf{1.14} & 1 & \cellcolor[gray]{0.9}4.69 & 3 \\
        \hline
        3 & cxxfilt$_3$ & \cellcolor[gray]{0.9}\textbf{3.41} & 3 & \cellcolor[gray]{0.9}162.62 & 31 \\
        \hline
        4 & giflib$_1$ & \cellcolor[gray]{0.9}\textbf{4.83} & 4 & \cellcolor[gray]{0.9}11.95 & 4 \\
        \hline
        5 & giflib$_2$ & \cellcolor[gray]{0.9}\textbf{15.87} & 11 & \cellcolor[gray]{0.9}111.60 & 21 \\
        \hline
        6 & jasper$_1$ & \cellcolor[gray]{0.9}\textbf{124.98} & 70 & \cellcolor[gray]{0.9}285.24 & 147 \\
        \hline
        7 & objdump$_1$ & \cellcolor[gray]{0.9}\textbf{27.37} & 18 & \cellcolor[gray]{0.9}142.06 & 64 \\
        \hline
        8 & libxml$_1$ & \cellcolor[gray]{0.9}\textbf{47.39} & 3 & \cellcolor[gray]{0.9}68.74 & 21 \\
        \hline
    \end{NiceTabular}
    \caption{Context weights experiments.}
    \label{tbl:exp_weights}
    \end{center}
\end{table}

\subsection{Time to Exposure}

Tables~\ref{tbl:tte_aflgo},~\ref{tbl:tte_beacon} show the results of TTE experiments. Columns \textbf{LibAFL-DiFuzz}, \textbf{AFLGo} and \textbf{BEACON} show time to exposure the crash for each instrument, respectively. Columns \textbf{Speedup} display the speedup of LibAFL-DiFuzz over each of the other instruments.

As we can see from the tables, our prototype managed to find crashes in \texttt{cxxfilt}, \texttt{giflib}, and \texttt{libxml} projects up to 93.78 times faster than AFLGo, and up to 86.38 times faster than BEACON. For \texttt{objdump} project LibAFL-DiFuzz managed to discover two distinct crashes while other instruments failed to find any.


\begin{table}[htbp]
    \begin{center}
    \small
    \begin{NiceTabular}{c | c | >{\columncolor[gray]{0.9}}c | c | >{\columncolor[gray]{0.9}}c}
        \hline
        № & TP ID & Time1 & Time2 & Speedup \\
        \hline
        1 & cxxfilt$_3$ & \textbf{3.41} & 56.43 & 16.55 \\
        \hline
        2 & cxxfilt$_4$ & \textbf{1.14} & 106.91 & 93.78 \\
        \hline
        3 & giflib$_1$ & \textbf{4.83} & 5.64 & 1.17 \\
        \hline
        4 & jasper$_1$ & 124.98 & \textbf{6.54} & 0.05 \\
        \hline
        5 & objdump$_2$ & \textbf{1.88} & - & - \\
        \hline
        6 & objdump$_3$ & \textbf{1.33} & - & - \\
        \hline
        7 & libxml$_1$ & \textbf{47.39} & 137.55 & 2.90 \\
        \hline
    \end{NiceTabular}
    \caption{TTE experiments LibAFL-DiFuzz vs. AFLGo.}
    \label{tbl:tte_aflgo}
    \end{center}
\end{table}

\begin{table}[htbp]
    \begin{center}
    \small
    \begin{NiceTabular}{c | c | >{\columncolor[gray]{0.9}}c | c | >{\columncolor[gray]{0.9}}c}
        \hline
        № & TP ID & Time1 & Time2 & Speedup \\
        \hline
        1 & cxxfilt$_3$ & \textbf{3.41} & 32.66 & 9.58 \\
        \hline
        2 & cxxfilt$_4$ & \textbf{1.14} & 98.47 & 86.38 \\
        \hline
        3 & giflib$_1$ & \textbf{4.83} & - & - \\
        \hline
        4 & jasper$_1$ & 124.98 & \textbf{11.29} & 0.09 \\
        \hline
        5 & objdump$_2$ & \textbf{1.88} & - & - \\
        \hline
        6 & objdump$_3$ & \textbf{1.33} & - & - \\
        \hline
        7 & libxml$_1$ & \textbf{47.39} & 199.93 & 4.22 \\
        \hline
    \end{NiceTabular}
    \caption{TTE experiments LibAFL-DiFuzz vs. BEACON.}
    \label{tbl:tte_beacon}
    \end{center}
\end{table}

However, for \texttt{jasper} our prototype shows worse results than AFLGo and BEACON. This may happen because of the small number of members in corresponding ETS and so complication of target approximation via proximity metrics.

Comparing to other instruments, LibAFL-DiFuzz has outperformed all of them on six out of seven chosen target points. The results of TTE experiments allow to conclude that LibAFL-DiFuzz mostly shows itself faster in crash reproduction.

\subsection{Total number of crashes}

Table~\ref{tbl:crashes} shows results of total crashes number experiments. Columns \textbf{LibAFL-DiFuzz}, \textbf{AFLGo}, and \textbf{BEACON} contain the total number of crash files found within specified time interval. 

\begin{table}[htbp]
    \begin{center}
    \small
    \begin{NiceTabular}{c | c | >{\columncolor[gray]{0.9}}c | c | c}
        \hline
        \textbf{№} & \textbf{Project} & \textbf{LibAFL-DiFuzz} &\textbf{AFLGo} &\textbf{BEACON} \\
        \hline\hline
        1 & cxxfilt & 10 & 164 & 114 \\
        \hline
        1 & giflib & 1399 & 14 & 11 \\
        \hline
        2 & jasper & $\approx$23k & 21 & 25 \\
        \hline
        3 & objdump & 254 & 0 & 0 \\
        \hline
        3 & libxml & 140 & 158 & 138 \\
        \hline
    \end{NiceTabular}
    \caption{Total crashes number experiments.}
    \label{tbl:crashes}
    \end{center}
\end{table}

As we see, the total amount of files provided by LibAFL-DiFuzz is greater than one provided by AFLGo and BEACON for \texttt{giflib}, \texttt{jasper}, and \texttt{objdump}, but less in remainder experiments. An important thing to notice is that most of found crash files cause the same crash, i.e. they are duplicates.

Another reason may be in the differences in architectural characteristics of tested instruments. Our prototype is based on LibAFL modular architecture, thus it spends some effort on communicating between its components and handling the necessary metadata. The overall speed of discovering crashes may be affected by these factors, giving different instruments some benefits in different experiments. However, LibAFL-based approach provides a lot of useful and convenient features that make an invaluable contribution to the convenience of fuzzer construction process, and form the universal platform for future fuzzing research.

\section{Conclusion}

We enhance directed fuzzing with context weights calculation for each target sequence component and implement it in our static analysis tool DiFuzz. We propose indirect edges resolving during call graph construction with the help of accurate handling of class hierarchies and their methods calls. Implementation of this technique is done in a separate library DiFuzzLLVM that handles LLVM IR of the tested program. We construct flexible, easy to configure architecture for directed fuzzing with components able to be easily combined with other techniques. Our prototype directed fuzzer called LibAFL-DiFuzz is based on LibAFL fuzzing library. 
We evaluated indirect calls resolving and got increase in directed fuzzing efficiency for reaching deeper target points.
We evaluated context weights contribution and got benefits in TTE and scheduling iterations number. We evaluated our fuzzer in comparison with AFLGo and BEACON, and got speedup in time to exposure up to 93.78 times. We also prevailed these tools in several usability characteristics important for the directed fuzzing process.

\section*{Future Work}

We highlight several directions for the future research in directed greybox fuzzing field and enhancing our method:
\begin{itemize}
    \item Improving performance of LibAFL-DiFuzz by enhancing LibAFL ETS-related components and their communication.
    \item Parallelization of adding static instrumentation to shorten time spent on this stage.
    \item Designing exploration-exploitation switching technique for balancing between opening new coverage and reaching target points.
    \item Integration LibAFL-DiFuzz with our symbolic execution tool Sydr~\cite{sydr} via incorporating into Sydr-Fuzz~\cite{sydr_fuzz} toolkit with their interaction through shared priority input queues.
\end{itemize}

\section{Acknowledgments}

We would like to thank Alexey Romanov from Lomonosov Moscow State University for greatly helpful consulting on LLVM.

\printbibliography


\end{document}